\documentclass[aps,prl,twocolumn,showpacs,floatfix,superscriptaddress,amsmath,amssymb]{revtex4-2}
\usepackage{amsfonts}
\usepackage{mathrsfs}
\usepackage{amsmath}
\usepackage{xcolor}
\usepackage{graphicx}
\usepackage{bm}
\usepackage{amssymb}
\usepackage{xspace}
\usepackage{epstopdf}
\usepackage{dcolumn}
\usepackage{longtable}
\usepackage{multirow}
\usepackage{float}
\usepackage{comment}
\usepackage{ulem} 
\usepackage[colorlinks=true, letterpaper=true, pdfstartview=FitV,  linkcolor=blue, citecolor=blue, urlcolor=blue]{hyperref}

\makeatletter

\begin{document}

\title{Design of altermagnetic models from spin clusters}

\author{Xingchuan Zhu}
\thanks{These authors contributed equally to this work.}
\affiliation{School of Intelligent Manufacturing, Nanjing University of Science and Technology, Nanjing 210094, China}

\author{Xingmin Huo}
\thanks{These authors contributed equally to this work.}
\affiliation{School of Physics, Beihang University,
Beijing 100191, China}

\author{Shiping Feng}
\affiliation{Faculty of Arts and Sciences, Beijing Normal University, Zhuhai 519087, China, and School of Physics and Astronomy,  Beijing Normal University, Beijing 100875, China}

\author{Song-Bo~Zhang}
\email{songbozhang@ustc.edu.cn}
\affiliation{Hefei National Laboratory, Hefei, Anhui 230088, China}
\affiliation{International Center for Quantum Design of Functional Materials (ICQD), University of Science and Technology of China, Hefei, Anhui 230026, China}

\author{Shengyuan A. Yang}
\affiliation{Research Laboratory for Quantum Materials, IAPME, FST, University of Macau, Taipa, Macau SAR, China}

\author{Huaiming Guo}
\email{hmguo@buaa.edu.cn}
\affiliation{School of Physics, Beihang University,
Beijing 100191, China}

\begin{abstract}
Altermagnetism, a new class of collinear compensated magnetic phase, has garnered tremendous interest because of its rich physics and promising applications. Physical models and verified material candidates for altermagnetism remain limited. 
Here, we propose a general scheme to construct altermagnetic models, which explicitly exhibits the blend of ferromagnetic and antiferromagnetic correlations in real space via the design of spin clusters, echoing the observation that properties of altermagnets resemble a mixture of ferromagnets and antiferromagnets. We show that in some of our models, the desired altermagnetic order can be spontaneously realized by electron-electron interaction in a broad range of the phase diagram.
This development facilitates the study of fascinating physics of altermagnetism and sheds light on the discovery of new
altermagnetic materials.

\end{abstract}


\maketitle


\textit{\textcolor{blue}{Introduction.--}}
Magnetic order, a central topic in condensed matter physics, is traditionally classified into ferromagnetism and antiferromagnetism.
Recently, a new class of collinear magnetic phases, termed altermagnetism, has been discovered, sharing aspects of both ferromagnetism and antiferromagnetism~\cite{Naka19NC,Ahn19PRB,Hayami20PRB,yuanLD20PRB,Smejkal2020,Mazin21PNAS,PhysRevX.12.031042,PhysRevX.12.040501,Bai2024Altermagnetism,jungwirth2024altermagnetsbeyondnodalmagneticallyordered}.
Like antiferromagnets, altermagnets feature fully compensated magnetization. 
However, in altermagnets, sublattices with opposite spins are linked by rotational or mirror symmetry (rather than inversion or translation)~\cite{PhysRevX.12.031042,PhysRevX.12.040501,PhysRevX.12.021016,PhysRevX.14.031037,PhysRevX.14.031038,PhysRevX.14.031039}. Consequently, 
energy bands of opposite spins are split even in absence of spin-orbit coupling, featuring characteristics of ferromagnets. Moreover, this spin splitting alternates signs in different regions of the Brillouin zone, as symmetry dictates.
These distinctive properties enable fascinating phenomena, such as anomalous Hall effect~\cite{Smejkal2020,feng2022anomalous}, spin transport without spin-orbit coupling~\cite{CJWu07PRB,Ma2021,PhysRevLett.126.127701,Bose22NE}, giant tunnelling magnetoresistance~\cite{Shao2021,PhysRevX.12.011028}, among others~\cite{YaoYG24PRL,guo2024valley,PhysRevB.110.184408,PhysRevLett.133.206702,DZhu24PRB,yan2024review,Reichlova2024,DuanXK25PRL,HJLin24arXiv,RChen25PRB}. In hybrid structures incorporating superconductors, more exotic phenomena were reported~\cite{Zhang2024,Ouassou23PRL,Papaj23PRB,PhysRevB.108.054511,Beenakker23PRB,PhysRevB.108.184505,PhysRevB.108.205410,PhysRevB.109.024517,Sumita23PRR,PhysRevB.110.L060508,hong2024unconventional,sim2024pairdensitywavessupercurrent,PhysRevB.108.205410,PhysRevLett.133.106601,PhysRevB.109.224502,PhysRevB.109.L201109,HaiP24arxiv}, including finite-momentum Cooper pairing~\cite{Zhang2024,hong2024unconventional,sim2024pairdensitywavessupercurrent,PhysRevB.110.L060508}, unconventional Andreev reflection~\cite{Papaj23PRB,PhysRevB.108.054511}, topological superconductivity~\cite{PhysRevB.108.205410,PhysRevLett.133.106601,PhysRevB.109.224502,PhysRevB.109.L201109,PhysRevB.108.184505}. These findings suggest great potential of altermagnets for spintronics and superconducting device applications.

\begin{figure*}[t]
  \includegraphics[width=1\linewidth]{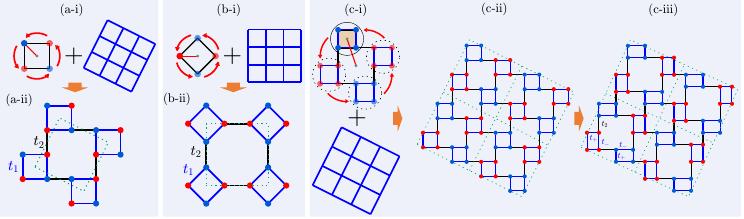}
  \caption{Schematics of realizing altermagnetism through spin cluster manipulation. (a-i), (b-i) and (c-i) depict the spin clusters and Bravais lattices used to construct altermagnets. Red (blue) dots represent sites with positive (negative) magnetic moments. (a-ii), (b-ii) and (c-ii) illustrate the resulting altermagnets on a $1/5$-depleted square lattice with antiferromagnetic plaquettes coupled ferromagnetically, a square-octagon lattice with antiferromagnetic plaquettes coupled ferromagnetically, and a $1/5$-depleted square lattice with ferromagnetic plaquettes coupled antiferromagnetically, respectively. (c-iii) Distorted clusters from square to rectangle shapes break $\mathcal{PT}$ symmetry emerging in (c-ii). All cases exhibit $d$-wave altermagnetism.} 
  \label{fig1}
\end{figure*}

Despite extensive research, so far, only a handful of {altermagnetic} candidates have been experimentally confirmed~\cite{krempasky2024altermagnetic,sciadv.adj4883,Lee24PRL,PhysRevLett.133.206401,ZengM2024AS,YangGW24arXiv,Reimers2024,Osumi24PRB,PhysRevLett.133.176401,Jiang2024arXiv,FaYZ24arxiv}.
Theoretical models for altermagnetism are also limited. Most rely on spin- and/or orientation-dependent hopping, which do not explicitly manifest the underlying magnetic order~\cite{PhysRevX.12.031042,PhysRevX.12.040501}.
In the few existing models that incorporate the magnetic order explicitly, the anisotropic spin splitting arises from the interplay between magnetic and non-magnetic sites or anisotropic local orbitals~\cite{PhysRevB.108.224421,PhysRevB.110.205140,chen2024spinexcitationsshastrysutherlandmodel,PhysRevB.108.L100402,antonenko2024mirrorchernbandsweyl,delre2024diracpointstopologicalphases,giuli2024altermagnetisminteractiondrivenitinerantmagnetism,PhysRevLett.132.236701,PhysRevLett.132.263402,kaushal2024altermagnetism}. It remains challenging to distinguish the magnetic order in altermagnets from that of conventional antiferromagnets.
To facilitate the study of altermagnetism, it is desirable to have more physical models discovered.

In this Letter, we propose a general scheme to generate altermagnetism. This method utilizes spin clusters as fundamental building blocks to generate the basis of the lattice model (Fig.~\ref{fig1}).
Within each cluster, the constituent spins are coupled either ferromagnetically or antiferromagnetically. These clusters are then arranged to form the lattice, which adheres to specific altermagnetic symmetries, such as $d$-, $g$-, and $i$-wave patterns.
Our approach naturally satisfies the requirements of broken spacetime inversion ($\mathcal{PT}$) symmetry and fully compensated magnetization. Importantly, the altermagnetic order is explicitly present in the model construction, showing a combination of
ferromagnetic and antiferromagnetic correlations. Using this approach, we generate concrete models for altermagnets with $d$-, $g$-, and $i$-wave symmetries. The resulting spin-splitting in the electronic spectrum and chirality-splitting in the magnonic spectrum are explicitly demonstrated. Furthermore, we show that in certain models, altermagnetism can be spontaneously realized through electron-electron interaction in a broad range of phase diagrams.
This scheme provides a versatile method for realizing altermagnetism across diverse geometries, significantly broadening the range of lattice models and material platforms available for exploring altermagnetic phenomena.

%

\textit{\textcolor{blue}{The general scheme.}} Our approach consists of the following steps: (i) Fix the symmetry of the model to be constructed. For altermagnetism, the key symmetry element is the one that connects the sublattices of opposite spins, denoted as $S$. Additional crystalline symmetries compatible with $S$ can also be imposed. (ii) Choose a Bravais lattice that is compatible with the prescribed symmetry. (iii) Design the basis. This is done by first choosing an arbitrary spin cluster. The basis is then generated from the spin cluster by imposing the symmetry $S$ and other crystalline symmetries. We put one orbital at each site of the basis, which for simplicity, we take to be an $s$-like orbital.
The sites are assigned with local spins, represented by red (upward) and blue (downward) colors.
The symmetry $S$ ensures that the number of red and blue sites is balanced within the basis.
(iv) Attach the basis to the Bravais lattice to form the lattice model. On each site with local spins, we add an on-site exchange field with signs determined by the color of the site. 
Spin-independent hoppings are introduced between neighboring sites.
(v) Check if the constructed model breaks the $\mathcal{PT}$ or $t\mathcal{T}$ symmetry ($t$ represents translation). If these symmetries are not broken, we can further distort the basis to achieve this. This step completes the construction of the altermagnetic model.

It is worth noting that in this approach, the prescribed symmetry is maintained at each Bravais lattice site in the basis.
Step (iii), it is helpful to require the basis to break the $\mathcal{PT}$ symmetry. Nevertheless, there may still be some accidental inversion centers in the lattice model, which makes the final check in step (v) necessary.

Compared to existing models, the models constructed in this scheme have the advantages that the altermagnetic order and its symmetry are made explicit via local spin configuration (colors of the sites), rather than relying on spin- and/or orientation-dependent hopping processes. The hopping terms here can be made as simple as possible, provided they fulfill the prescribed symmetry. Furthermore, the construction allows for considerable freedom in choosing both the lattice geometry and local spin configuration.

\textit{\textcolor{blue}{Example: 2D model with $d$-wave altermagnetism.--}}We demonstrate our scheme by constructing a two-dimensional (2D) model for $d$-wave altermagnetism.
To realize $d$-wave type spin splitting, the key symmetry element is $S=C_{4z}\mathcal{T}$. 
The 2D Bravais lattice compatible with this symmetry is the square lattice. These finish the first two steps of our procedure. 

For the basis construction, we choose a single magnetized site for each spin cluster, which is the simplest possible case. 
Assume this site, labeled as $b_1$, has a local up-spin. Clearly, $b_1$ alone cannot satisfy $S=C_{4z}\mathcal{T}$. To generate a basis respecting $S$, we let $S$ operate on $b_1$ and generate three additional sites $b_i$ ($i=2,3,4$), as shown in Fig.~\ref{fig1}(a-i). This results in a basis consisting of four sites. Due to the constraint of symmetry $S$, the local spins are fully compensated in the basis: there are two red sites and two blue sites. Moreover, the basis breaks the $\mathcal{PT}$ symmetry while preserving inversion symmetry.

Next, we attach the basis to the square Bravais lattice and construct the model Hamiltonian by adding on-site exchange and hopping terms. There is freedom in choosing the orientation of the basis with respect to the Bravais lattice. 
In the current case, for example, we obtain two different geometries, shown in Figs.~\ref{fig1}(a-ii) and \ref{fig1}(b-ii), the 1/5-depleted square lattice and the square-octagon lattice, respectively. 
If only nearest-neighbor hopping processes are included, these two geometries lead to equivalent models. Their differences emerge when further-neighbor hopping processes are added. For simplicity, we focus on the case with nearest-neighbor hopping. As mentioned earlier, it is sufficient to consider only spin- and orientation-independent hopping, as illustrated in Fig.~\ref{fig1}(a-ii) and \ref{fig1}(b-ii). In the final step, we confirm that the model does not have  $\mathcal{PT}$ and any $t\mathcal{T}$ symmetry. This completes the construction.

%
%
%

\begin{figure}[t]
  \includegraphics[width=1\linewidth]{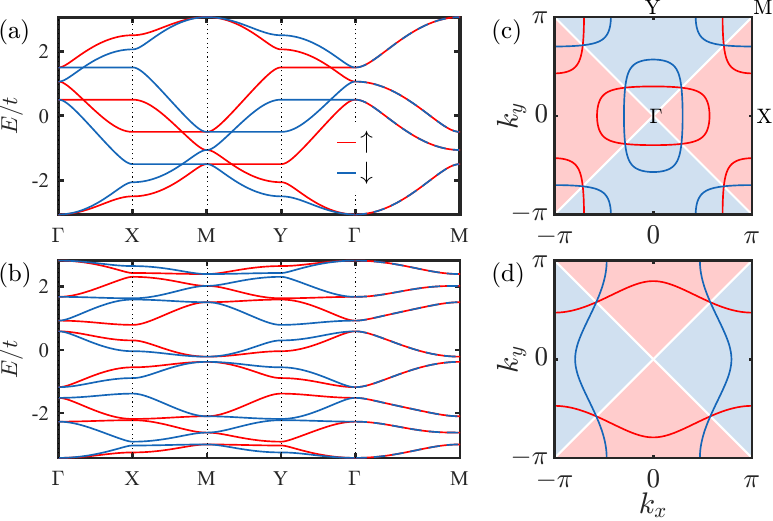}
  \caption{(a) Band structure of the $1/5$-depleted square lattice with intracell antiferromagnetic order along high symmetry lines in the BZ. Red (blue) lines represent the spin-up (spin-down) bands.
  (b) Same as (a), but for the 1/5-depleted square lattice with intracell ferromagnetic order. (c) and (d) show Fermi surfaces corresponding to (a) and (b), respectively. In (a), the parameters are $t_1=t_2=1$, ${JS}=0.5$ and $\mu = 0$, while in (b), $t_{+} = 1.6$, $t_{-} = 0.4$, $t_2=1$, ${JS}=0.5$ and $\mu = 0.3$. }
  \label{fig2}
\end{figure}

The electronic tight-binding model for this example can be written as 
\begin{align}
H=&-t_1 \sum_{\langle i j\rangle_1, \sigma} c_{i, \sigma}^{\dagger} c_{j, \sigma}-t_2 \sum_{\langle i j\rangle_2, \sigma} c_{i, \sigma}^{\dagger} c_{j, \sigma} \nonumber
\\
&-J \sum_{i, \sigma, \sigma^{\prime}} \boldsymbol{S}_i \cdot c_{i, \sigma}^{\dagger} \boldsymbol{\sigma}_{\sigma \sigma^{\prime}} c_{i, \sigma^{\prime}}-\mu \sum_{i, \sigma} c_{i, \sigma}^{\dagger} c_{i, \sigma}.
\label{eq1}
\end{align}
Here, $c_{i,\sigma}^{\dagger}$ and $c_{i,\sigma}$ are fermionic creation and annihilation operators with spin $\sigma\in\{\uparrow,\downarrow\}$ at site $i$. The notation $\langle ij \rangle_{p}$ represents nearest-neighbor pairs within a basis (intracell) for $p=1$, and pairs between neighboring basis (intercell) for $p=2$. The $J$ term describes the on-site exchange interaction between the local spin $\bm{S}_i$ and itinerant electron spin $
\bm{\sigma}$. The chemical potential $\mu$ determines the electron filling of the system. {In the calculations, the $J$ term is treated using a mean-field approximation with the fluctuations of the local spin neglected, and the frozen magnetic moments adopt a configuration constructed in the above scheme.}

In the absence of magnetic order ($J=0$), the band structure consists of four spin-degenerate bands. Upon incorporating the altermagnetic order ($J\neq 0$), the spin degeneracy is lifted, despite vanishing net magnetization. This is evidenced in the calculated band structure [Fig.~\ref{fig2}(a)]. Due to the $C_{4z}\mathcal{T}$ symmetry, the spin splitting exhibits a $d$-wave character in momentum space. {In Fig.~\ref{fig2}(c)}, we present the spin-resolved Fermi surfaces for $\mu=0$, which further confirms this feature of altermagnetism:  There are two pairs of Fermi surfaces, centered at $\Gamma$ and $M$ points, respectively. The spin-up and spin-down Fermi surfaces display significant anisotropy and are related to each other by a $\pi/4$ rotation.

In the above construction, we use a single site to generate the basis, which is the simplest choice. More generally, a spin cluster can be chosen for this purpose. For instance, we take four magnetic sites with the same spin (depicted in red) as the cluster and generate a basis by applying $S=C_{4z}\mathcal{T}$. This process yields a configuration consisting of 16 sites [Fig.~\ref{fig1}(c-i)], leading to the lattice model illustrated in Fig.~\ref{fig1}(c-ii). Notably, this new model can be viewed as a modification of the model in Fig.~\ref{fig1}(a-i), where each site is replaced by a spin cluster of four sites with the same spin. This modification introduces a $\mathcal{PT}$ symmetry with an inversion center located between two neighboring spin clusters [Fig.~\ref{fig1}(c-ii)]. Nevertheless, breaking this symmetry is straightforward, for example, by distorting the cluster from a square to a rectangle, as shown in Fig.~\ref{fig1}(c-iii). The calculated band structure and Fermi surfaces in Fig.~\ref{fig2} confirm the $d$-wave altermagnetic character.

\begin{figure}[t]
  \includegraphics[width=1.0\linewidth]{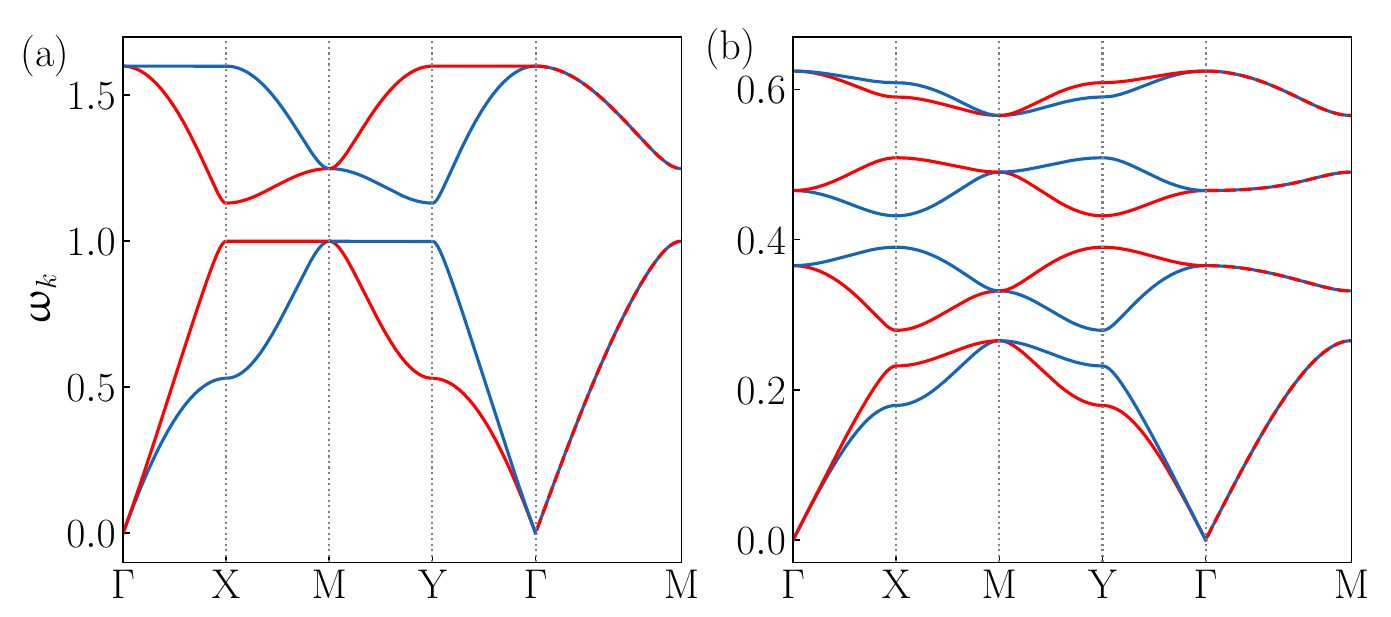}
  \caption{Magnon dispersion, $\omega_{\bf k}$, in the $1/5$-depleted square lattice: (a) intracell AF order; (b) intracell FM order, corresponding to Fig.~\ref{fig1} (a-ii) and (c-iii), respectively. In (a), the parameters are $J_1=1, J_2=0.6$, while in (b), $J_{+}=0.2, J_{-}=0.1$, and $J_2=1$, respectively.}
  \label{fig3}
\end{figure}

\textit{\textcolor{blue}{Chirality splitting in magnon spectrum.--}}
With explicit magnetic orders, the constructed models also provide a straightforward way to explore another key feature of altermagnetism: chirality splitting in the magnon spectrum~\cite{PhysRevLett.131.256703,PhysRevLett.133.156702}. 
To investigate this phenomenon, we begin by writing the spin Hamiltonian for the given model. As an example, consider the model shown in Fig.~\ref{fig1}(a-ii). The Heisenberg spin Hamiltonian for this model is given by
\begin{eqnarray}
H_m = J_1 \sum_{\langle ij \rangle_1} {\bf S}_i\cdot {\bf S}_j - J_2 \sum_{\langle ij \rangle_2} {\bf S}_i\cdot {\bf S}_j,
\end{eqnarray}
where the $J_1$ and $J_2$ terms represent the intracell (antiferromagnetic) and intercell (ferromagnetic) exchange couplings, respectively.
The ground state of this Hamiltonian corresponds to the $Q = (0,0)$ antiferromagnetic order. To study magnons, we employ linear spin-wave theory and use the Holstein-Primakoff transformation, which expresses the spin operators in terms of bosonic operators $a$ and $a^\dagger$. For spin-up sites ($i=1,3$), we put $S_{i}^x+iS_{i}^y = \sqrt{2S} a_{i}$, $S_{i}^x-iS_{i}^y = \sqrt{2S} a_{i}^\dagger$ and $S_{i}^z = S - a_{i}^\dagger a_{i}$; whereas for spin-down sites ($i=2,4$), $S_{i}^x+iS_{i}^y = \sqrt{2S} a_{i}^\dagger$, $S_{i}^x-iS_{i}^y = \sqrt{2S} a_{i}$ and $ S_{i}^z = a_{i}^\dagger a_{i} - S$. After Fourier transform and in the basis $\Psi_{\bf k}=(a_{{\bf k}, 1}, a_{-{\bf k}, 2}^{\dagger}, a_{{\bf k}, 3}, a_{-{\bf k}, 4}^{\dagger})^T$, the Hamiltonian can be rewritten as
\begin{eqnarray}
{\cal H}_m({\bf k})= - \dfrac{1}{2}\left[\begin{array}{cccc}
0 & -J_1 & {J_2}e^{-ik_y}  & -J_1 \\
-J_1 & 0 & -J_1  & J_2e^{-ik_x}  \\
{J_2}e^{ik_y} & -J_1  & 0 & -J_1  \\
-J_1 & {J_2}e^{ik_x} & -J_1 & 0
\end{array}\right],
\end{eqnarray}
where we have dropped a constant term $J_1+J_2/2$. The resulting magnon spectrum is presented in Fig.~\ref{fig3}.
The red (blue) color indicates that the mode has a dominant amplitude on spin-up (down) sites and exhibits positive (negative) chirality. This confirms the chirality splitting, which follows the $d$-wave pattern, similar to the spin-splitting observed in the electronic spectrum. Moreover, around the $\Gamma$ point (${\bf k}=0$), the lowest magnon branch exhibits a linear dispersion given by $\omega=\frac{\sqrt{5}}{2} \frac{J_1 \sqrt{J_2}}{\sqrt{J_1+J_2}} k$, a characteristic typically observed in antiferromagnets.

\begin{figure}[t]
  \includegraphics[width=1\linewidth]{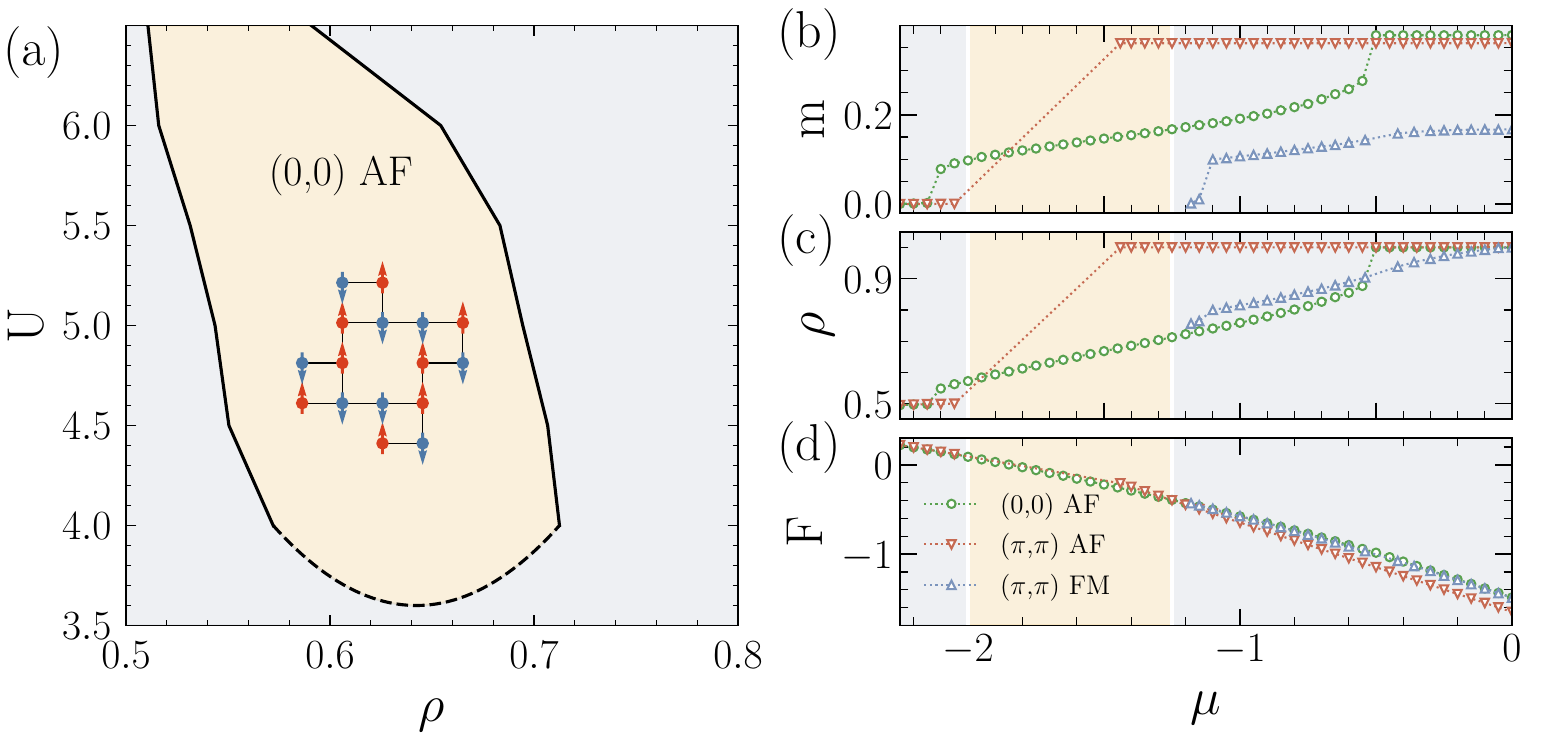}
  \caption{(a) Phase diagram in $\rho$-$U$ parameter space. The yellow region indicates the (0,0) antiferromagnetic phase. (b) Order parameter $m$, (c) average density $\rho$ and (d) free energy $F$ as functions of chemical potential for various magnetic orders on the $1/5$-depleted square lattice. The parameters are $t_1=t_2=1$, and $U=4$.}
\label{fig4}
\end{figure}

\textit{\textcolor{blue}{Interaction-driven altermagnetism.--}} The altermagnetic order in the models constructed here may spontaneously emerge from electron-electron interactions. To illustrate this, let us focus on the 1/5-depleted square lattice depicted in Fig.~\ref{fig1}(a-i) and consider on-site Hubbard interaction
$U \sum_i n_{i \uparrow} n_{i \downarrow}$, where $n_{i\uparrow(\downarrow)}$ represents the spin-dependent charge density operator. At half-filling, the dominant magnetic order corresponds to the conventional N\'eel order, as confirmed by the determinant quantum Monte Carlo (DQMC) method~\cite{PhysRevLett.113.106402}.
Away from half-filling, the DQMC method is hindered by the minus-sign problem. {{Hence, we employ a mean-field analysis, a widely used approach for studying magnetically ordered phases~\cite{PhysRevB.108.L100402,PhysRevLett.132.263402,PhysRevB.31.4403,SSorella_1992,Claveau_2014}. Within this approximation, the Hubbard term is decomposed as}} $n_{i \uparrow} n_{i \downarrow} \approx n_{i \uparrow}\left\langle n_{i \downarrow}\right\rangle+\left\langle n_{i \uparrow}\right\rangle n_{i \downarrow}-\left\langle n_{i \uparrow}\right\rangle\left\langle n_{i \downarrow}\right\rangle$. The spin-dependent density is parameterized as $\left\langle n_{i \uparrow(\downarrow)}\right\rangle=\rho/2\pm m$, where $\rho$ is the average density and $m$ the order parameter. At zero temperature, the free energy is computed by summing the negative eigenvalues of the resulting mean-field Hamiltonian: $F=\sum_{E_i<0}E_i+E_0$, where $E_0$ is a constant energy (detailed in Supplemental Material~\cite{SM}). The order parameter $m$ is determined self-consistently by minimizing the free energy.

We calculate three distinct magnetic orders: $(\pi,\pi)$ antiferromagnetic, $(0,0)$ antiferromagnetic and $(\pi,\pi)$ ferromagnetic {orders} on the lattice. The phase diagram against the filling $\rho$ and interaction strength $U$ is shown in Fig.~\ref{fig4}(a). Strikingly, the $(0,0)$ antiferromagnetic order can be stabilized across a broad region near the filling $\rho=0.6$ and interaction strength $U=5$. In this region, the $(0,0)$ antiferromagnetic order emerges as the sole self-consistent solution or displays the lowest free energy compared to other magnetic configurations and the absence of any magnetic order [Figs.~\ref{fig4}(b-d)].

\begin{figure}[t]
  \includegraphics[width=0.9\linewidth]{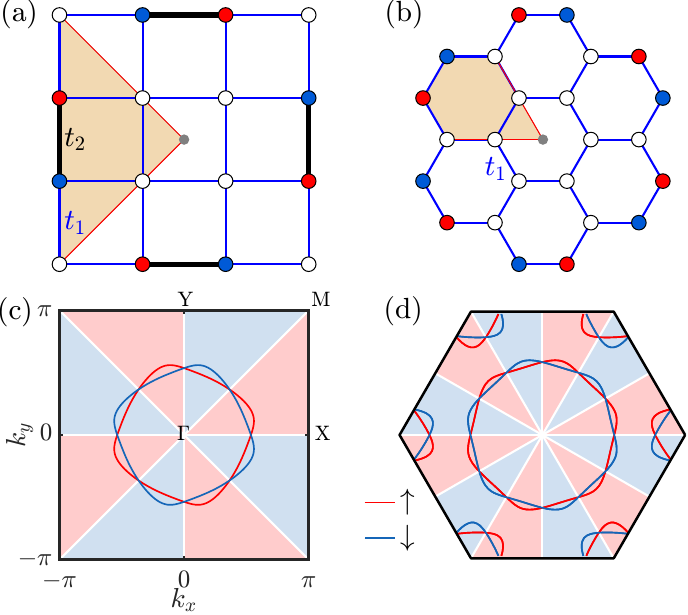}
  \caption{(a) Spin cluster in a square geometry to realize $g$-wave altermagnetism. (b) Spin cluster in a honeycomb geometry to realize $i$-wave altermagnetism. (c,d) Fermi surfaces of the altermagnetic lattice models based on the spin clusters in (a,b), respectively. White sites in (a,b) indicate zero magnetic moments, while other colored sites are the same with those in Fig.~\ref{fig1}. The parameters for (c) are $t_1 = 1$, $t_2 = 0.4$, $\mu = 1.5$ and $JS = 0.8$, and for (d), $t_1 = 1$, $\mu = 1.3$ and $JS = 0.9$.}
\label{fig5}
\end{figure}

\textit{\textcolor{blue}{Discussion.--}}
The altermagnetic models constructed using our scheme explicitly display a mixture of antiferromagnetic and ferromagnetic correlations, making it intuitive to understand that altermagnetism blends the physical properties of both antiferromagnets and ferromagnets. This scheme provides a simple yet versatile approach for designing altermagnetism with any tailored symmetries and geometries. In Fig.~\ref{fig5}(a,b), we exemplify $g$-wave and $i$-wave altermagnetism in a square and a honeycomb lattice, respectively. In these models, the symmetries are directly achieved by real-space magnetic orders, which ensures the corresponding symmetry of spin splitting in momentum space, as shown in Figs.~\ref{fig5}(c,d). Additional examples can be found in Supplemental Material~\cite{SM}, where we also demonstrate the applicability of this scheme to three dimensions. 

Our proposed models offer a concrete platform for exploring rich physics associated with altermagnetism and provide valuable guidance for searching and designing altermagnetic materials. In particular, the $1/5$-depleted square lattice shown in Fig.~\ref{fig1}(a) is a recurring structure in magnetic compounds such as octagraphene\cite{Li_2022}, CaV$_4$O$_9$~\cite{doi:10.1143/JPSJ.64.2758}, NaKV$_4$O$_9\cdot$H$_2$O~\cite{C5DT04745E}, K$_{0.8}$Fe$_{1.6}$Se$_2$~\cite{Bao_2011}, and A${_2}$Fe${_4}$Se${_5}$ (A =K, Rb, Cs, TI)~\cite{PhysRevLett.107.137003}. This suggests that these systems could serve as promising platforms to explore altermagnetism.

\begin{acknowledgments}
We thank Z.-B. Yan, L.-H. Hu, Z.-Q. Liu, Y. Du, C.-C. Liu, X.-L. Sheng, Y.-J. Zhao, and D. L. Deng for
helpful discussions. X.H. and H.G. acknowledge support from the NSFC grant No.~12074022 and the BNLCMP open research fund under Grant No.~2024BNLCMPKF023. X.Z. acknowledges support from the Natural Science Foundation of Jiangsu Province under Grant BK20230907 and the NSFC grant No.~12304177. S.F. is supported by the National Key Research and Development Program of China under Grant Nos. 2023YFA1406500 and 2021YFA1401803, and NSFC under Grant No. 12274036. S.Y. acknowledges support from UM MYRG (GRG2024-00018-IAPME). S.B.Z. was supported by the start-up fund at HFNL, the Innovation Program for Quantum Science and Technology (Grant No. 2021ZD0302800).
\end{acknowledgments}

\bibliographystyle{apsrev4-2}
\bibliography{ref}

\end{document}